\documentstyle[12pt]{article}
\begin{document}
\begin{flushright}
PRA-HEP-93/07\\
June 1993
\end{flushright}
\vspace{5ex}
\begin{center}
{\Large {\bf Direct vs. resolved photon: an exercise in
factorization }} \\
\vspace{0.5cm}
{Ji\v{r}\'{\i} Ch\'{y}la \\
\vspace{0.4cm}
Institute of Physics, AV\v{C}R,
Na Slovance 2, Prague 8, Czech Republic\\}
\end{center}
\vspace{0.5cm}
{\bf Abstract}
Direct and resolved photon interactions are shown to be intimately
related through the factorization mechanism. It is argued that in
theoretically consistent analysis of jet production in $\gamma$p
and ep collisions the LO resolved $\gamma$ contribution must be
considered together with the NLO direct $\gamma$ component.
Recent data from HERA therefore do
not provide a direct evidence for the former component, but should
rather be interpreted as a manifestation of the
$O(\alpha^2\alpha_s^2)$ term in ep interactions.

\vspace{1.0cm}
In recent months an increasing flow of fresh data
from HERA at DESY has brought several interesting new results
on the structure of photon. Among them, the interpretation
\cite{H1,Zeus} of data on two-jet events
an as evidence for the resolved $\gamma$ contribution to $\gamma$p
interactions has attracted considerable attention. I consider this
interpretation as premature. Due to close relation
between direct and resolved $\gamma$ components, based on the
general factorization arguments, there is in fact no principal
difference between these two contributions.
I shall argue that in theoretically consistent
treatment of $\gamma$p interactions, the usual resolved $\gamma$
component should always be considered simultaneously with the NLO
direct one, which, in fact, leads to kinematically very similar final
state configurations. For the case of quarks
inside the photon, the separation of $\gamma$p interactions into
``direct'' and ``resolved'' components is theoretically ambiguous and
doesn't allow a unique interpretation  of observed jet correlations
as an evidence for the ``resolved'' photon.

Let me start with an important comment on the photon structure
functions.
To the order $O(\alpha\alpha_s)$ the evolution equations for quark
(of flavour $i$ and charge $e_i$),
gluon and photon distribution functions inside the real photon read
\begin{equation}
\frac{{\rm d}D_{\gamma/\gamma}(x,M)}{{\rm d}\ln M}= 0, \;\;\;\;\;
D_{\gamma/\gamma}(x,M)=\delta(1-x)
\label{gamma}
\end{equation}
\begin{equation}
\frac{{\rm d}D_{g/\gamma}(x,M)}{{\rm d}\ln M}=
\frac{\alpha_s}{\pi} \sum_{j=1}^{2n_f} \int^1_{x}
\frac{{\rm d}y}{y} D_{q_j/\gamma}
(y,M)P_{g/q}^{(0)}\left(\frac{x}{y}\right)+
\frac{\alpha_s}{\pi} \int^1_{x}\frac{{\rm d}y}{y} D_{g/\gamma}
(y,M)P_{g/g}^{(0)}\left(\frac{x}{y}\right)
\label{gluon}
\end{equation}
\begin{equation}
\frac{{\rm d}D_{q_i/\gamma}(x,M)}{{\rm d}\ln M}=
\frac{\alpha}{\pi}k_i(x)+
\frac{\alpha_s}{\pi} \int^1_{x}\frac{{\rm d}y}{y} D_{q_i/\gamma}
(y,M)P_{q/q}^{(0)}\left(\frac{x}{y}\right)+
\frac{\alpha_s}{\pi} \int^1_{x}\frac{{\rm d}y}{y} D_{g/\gamma}
(y,M)P_{q/g}^{(0)}\left(\frac{x}{y}\right)
\label{quark}
\end{equation}
where $k_i(x)=3e_i^2(x^2+(1-x)^2)$ is the LO QED branching
function corresponding to the vertex in Fig.1a and $P^{(0)}_{a/b}(z)$
are the analogous LO QCD branching functions. In
(\ref{gamma}-\ref{quark}) the
argument (not written out explicitly) of the strong coupling constant
$\alpha_s$,
the so-called factorization scale $M$, is
the same as the scale of the various distribution functions.
In the above equation the term ``quarks'' denotes
both quarks and antiquarks
and correspondingly the summation in (\ref{gluon}) runs
from 1 to $2n_f$.

In many papers, e.g. [3-7],
dealing with the photon distribution functions
 one finds the claim that the
resolved $\gamma$ contribution, described by Feynman diagrams
like those of Fig.2a, is of the order $O(\alpha\alpha_s)$,
despite the presence of two strong interaction vertices.
This claim is based on
incorrect analysis of the behaviour of $D_{q/\gamma}(x,M)$ in the limit
$\alpha_s\rightarrow0$, which has lead to the wrong conclusion that
it behaves like $O(\alpha/\alpha_s)$, the $\alpha_s$ in the denominator
cancelling one power of $\alpha_s$ from the vertices.
To show this, let us investigate, in LO QCD,
the behaviour of $D_{q/\gamma}$ in weak coupling region.
In the simplest case of the
generic nonsinglet quark distribution function, $D_{\rm NS}(x,M)$,
the appropriate evolution equation reads \cite{Gluck}
\begin{equation}
\frac{{\rm d}D_{\rm NS}(x,M)}{{\rm d}\ln M}=
\frac{\alpha}{\pi} k_{\rm NS}(x) + \frac{\alpha_s}{\pi}
\int^1_{x}\frac{{\rm d}y}{y} D_{\rm NS}
(y,M)P_{q/q}^{(0)}\left(\frac{x}{y}\right),
\label{NSAPQ}
\end{equation}
where
\begin{equation}
\alpha_s(M)=\frac{4\pi}{\beta_0\ln(M^2/\Lambda^2_{\rm LO})}.
\label{alphalo}
\end{equation}
Switching off QCD already at this stage, i.e. setting $\alpha_s=0$ on
the r.h.s. of (\ref{NSAPQ}), we get
\begin{equation}
D_{\rm NS}^{\rm QED}(x,M)=
\frac{\alpha}{\pi}k_{\rm NS}(x)\ln \frac{M}{M_0},
\label{simple}
\end{equation}
where $M_0$ is an arbitrary positive constant ($M$ is by definition
positive).  Eq. (\ref{simple}) is, in fact, what we get by integrating
the pole part
\begin{equation}
\frac{\alpha}{2\pi}k_{\rm NS}(x)\frac{1}{-t}
\label{polepart}
\end{equation}
of the corresponding
Feynman diagrams in Fig.2b, in the interval $\mid t\mid\in(M_0^2,M^2)$.

The claim that $D_{\rm NS}(x,M)={\cal O}(\alpha/\alpha_s)$ or, in
general, $D_{q/\gamma}(x,M)={\cal O}(\alpha/\alpha_s)$, is based on
the existence of the so-called  ``asymptotic pointlike'' (or
``anomalous'') solution to the full evolution equation (\ref{NSAPQ}),
which is explicitly calculable in perturbative QCD
\begin{equation}
D_{\rm NS}^{ap}(x,M)=\frac{4\pi}{\alpha_s(M)}a_{\rm NS}(x),
\label{asymptotic}
\end{equation}
where $a_{\rm NS}(x)$ is given as a solution to the equation
\begin{equation}
a_{\rm NS}(x)=\frac{\alpha}{2\pi\beta_0}k_{\rm NS}(x)+\frac{2}{\beta_0}
\int_{x}^{1}\frac{{\rm d}y}{y}P_{q/q}^{(0)}\left(\frac{x}{y}\right)
a_{\rm NS}(y),
\label{a(x)}
\end{equation}
and which indeed behaves like $O(\alpha/\alpha_s)$!
However, as we shall see the existence of this solution provides no
justification for the claim that the full solution of (\ref{NSAPQ})
behaves like ${\cal O}(\alpha/\alpha_s)$.
Let me first recall the important fact \cite{Gluck}
that this asymptotic pointlike solution doesn't
exist alone but is, together with the solution of the
corresponding homogeneous equation (describing the ``hadronic'' part of
the photon), embedded in the general solution of the
evolution equation (\ref{NSAPQ}).  Converting both these components
into moments for easier handling, we get for it the expression
\cite{Gluck}:
\begin{equation}
D_{\rm NS}(n,M)=\frac{4\pi}{\alpha_s(M)}\left[1-\left(\frac{\alpha_s(M)}
{\alpha_s(M_0)}\right)^{1-2P^{(0)}(n)/\beta_0}\right]a_{\rm NS}(n)+
\left[\frac{\alpha_s(M)}{\alpha_s(M_0)}\right]^{-2P^{(0)}(n)/\beta_0}
D_{\rm NS}(n,M_0)
\label{full solution}
\end{equation}
which has been obtained as the sum of the asymptotic pointlike solution
(\ref{asymptotic}) and the general solution to the
corresponding homogeneous equation
\begin{equation}
D_{\rm NS}^{had}(n,M)=D_{\rm NS}^{had}(n,M_0)
\left[\frac{\alpha_s(M)}{\alpha_s(M_0)}
\right]^{-2P^{(0)}(n)/\beta_0},
\label{hadronic}
\end{equation}
taking into account (\ref{a(x)}). $M_0$ in the above equations denotes
the scale at which the boundary condition on $D_{\rm NS}(x,M_0)$ is
specified.

 In the weak coupling limit, i.e. for
$\Lambda_{\rm QCD}\rightarrow0$, the second term in
(\ref{full solution}) goes over into $M$-independent
$D_{\rm NS}(x,M_0)$, i.e. just the
initial condition, while the dominant first term behaves as
\begin{equation}
 \frac{4\pi}{\alpha_s(M)}\frac{(1-2P^{(0)}(n)/\beta_0)\ln(M/M_0)} {\ln
(M/\Lambda_{\rm QCD})}a_{\rm NS}(n)\rightarrow
\left(1-2P^{(0)}(n)/\beta_0\right)\frac{\alpha}{\pi}k_{\rm NS}(n)
\ln \frac{M}{M_0}
\label{limit}
\end{equation}
and is thus very close to the simple expression (\ref{simple}),
expected solely on the
ground of QED vertex in Fig.1, with no QCD effects at all. The
only difference with respect to (\ref{simple})
rests in the numerical factor
$(1-2P^{(0)}(n)/\beta_0)$
containing the term proportional to $P^{(0)}(n)$, which
is a consequence of the LO QCD corrections (like those of Fig.1b) to
the basic QED vertex of Fig.1a. As the limit (\ref{limit})
is, as noticed already in \cite{Field}, finite, there is no
justification to write $D_{\rm NS}(x,M)={\cal O}(\alpha/\alpha_s)$. The
same is of course true for the singlet quark distribution function and
for $D_{g/\gamma}$ as well.  The preceding considerations clearly show
how important it is to consider the asymptotic pointlike solution always
in conjunction with the hadronic component of the photon.
Recall, as another example, that $D_{q/\gamma}^{ap}(x,M)$ in
(\ref{asymptotic}) diverges badly at $x=0$. Also in this
case the problem is cured \cite{Gluck} in the sum (\ref{full solution}).

The crucial aspect of the above procedure is obviously the way
the weak coupling limit is constructed.
Let us compare the three above mentioned
expressions (\ref{simple},\ref{asymptotic},\ref{full solution})
for $D_{\rm NS}(x,M)$ in this limit,
obtained by sending $\Lambda_{\rm QCD}\rightarrow0$.
They differ in two aspects: the place where
the limitting procedure was carried out and the interplay
with the hadronic component.
In the case of (\ref{simple}), $\alpha_s$ was set to zero in the
evolution equation itself, leading to a well-defined, finite result
for all $M$, which, however, carries no trace of the QCD corrections,
present for $\alpha_s\neq0$.
For the asymptotic pointlike solution (\ref{asymptotic}),
considered separately from the hadronic part (\ref{hadronic}),
the limit $\Lambda_{\rm QCD}\rightarrow0$ leads
to the result $D^{ap}_{\rm NS}(x,M)\rightarrow\infty$ for
any $x, M$. The hadronic component (\ref{hadronic}), on the other hand,
approaches a finite limit, given, as
expected, by the initial distribution function
$D^{had}_{\rm NS}(x,M_0)$!  This is another signal that $D^{ap}_{\rm
NS}(x,M)$ has little physical meaning of its own but makes sense only
in combination with the hadronic part of the photon. In other words,
taking first the limit $\Lambda_{\rm QCD}\rightarrow0$ and then adding
the two components leads to ill-defined result.

If, on the other hand, we first add in (\ref{full solution}) the two
mentioned components and only then take the limit
$\Lambda_{\rm QCD}\rightarrow0$ we get not only a well-behaved result
(\ref{limit}), but this finite result contains (in the term proportional
to $P^{(0)}(n)$) also a clear trace of the presence of higher order QCD
corrections. Note that this finite limit, coming
from the first term on the r.h.s. of (\ref{full solution}), is actually
independent of the parametrization of the hadronic part
$D^{had}_{\rm NS}(x,M)$.

In summary, the photon distribution function $D_{q/\gamma}$
is clearly of the order ${\cal O}(\alpha)$ and consequently the
resolved photon contribution to, for instance, jet
production in $\gamma$p collisions of the order
${\cal O}(\alpha\alpha_s^2)$ and not, as claimed in [3-7],
of the order ${\cal O}(\alpha\alpha_s)$. It should
therefore be added to the NLO direct $\gamma$ contribution,
which is of the same order.

The close relation between the LO resolved and NLO direct $\gamma$
contributions is in fact central to the very idea of factorization
of parallel singularities.
It also turns out that the factorization of parallel
singularities provides another
clear evidence that $D_{q/\gamma}(x,M)={\cal
O}(\alpha)$.
To demonstrate this assertion, let me
discuss, at the NLO, the factorization of parallel
singularities into the structure functions of beam particles in three
distinct cases:
\begin{equation}
{\rm p + p}\rightarrow {\rm 2\; jets + anything}
\label{pp}
\end{equation}
\begin{equation}
\gamma+{\rm p} \rightarrow {\rm 2\; jets + anything}
\label{photoproduction}
\end{equation}
\begin{equation}
\gamma+{\rm p} \rightarrow W + {\rm anything}
\label{W}
\end{equation}
Consider the basic QED vertex of Fig.1. Assuming that it
is the lower, quark, leg which enters the interaction vertex in Fig.2b,
we encounter a singularity of the form $1/t$. Its factorization then
amounts to dividing the whole integration range into the lower part
$\mid t\mid\in(t_{min},M^2)$ ($t_{min}$ is essentially an infrared
regulator), put into the quark distribution function
$D_{q/\gamma}(x,M)$ and used in the LO resolved $\gamma$ contribution of
Fig.2a, and the upper one, $\mid t\mid\in(M^2,t_{max})$, retained in the
NLO hard scattering cross-section of the direct $\gamma$ subprocess of
Fig.2b.
It seems obvious that due to the arbitrariness in the choice of the
boundary value $M^2$ between what is included in the ``resolved'' and
what is left in ``direct'' $\gamma$ component, these components have to
be of the same order and thus $D_{q/\gamma}(x,M)$ of the order
$O(\alpha)$.

The general idea of factorization of parallel singularities
\cite{Politzer} implies the following structure of the
cross-section for the process (\ref{pp}), described by Feynman diagrams
in Fig.3
\footnote{In Figures 3-5 only examples of typical Feynman diagrams,
relevant to the discussed point, are shown.}
\begin{equation}
\sigma({\rm pp}\rightarrow {\rm 2\; jets})=
\sum_{a,b}\int_0^1{\rm d}x_1\int_0^1{\rm d}x_2
D_{a/p}(x_1,M)D_{b/p}(x_2,M)
\sigma^{hard}_{ab}(S,x_1,x_2,M,\mu),
\label{sigmahadrons}
\end{equation}
where the hard scattering cross-section $\sigma^{hard}_{ab}$
for the parton level subprocess
$a+b\rightarrow c+d+\cdots$
admits perturbation expansion
in powers of $\alpha_s(\mu)$ at the
                 hard scattering scale $\mu$,
generally different from the factorization scale $M$:
\begin{equation}
\sigma^{hard}_{ab}(S,x_1,x_2,M,\mu)=
\left[\alpha_s^2(\mu)\kappa^{\rm LO}_{ab}(S,x_1,x_2)+
\alpha_s^3(\mu)\kappa^{\rm NLO}_{ab}(S,x_1,x_2,M,\mu)\right],
\label{hardhadrons}
\end{equation}
where $S$ is the total pp CMS energy, the sum in (\ref{sigmahadrons})
runs over all parton pairs $a,b$ in the incoming hadrons.
The functions $\kappa^{\rm LO}, \kappa^{\rm NLO}$ are assumed to
contain all appropriate $\delta$-functions defining the kinematics
of the final state jets in (\ref{pp}).
Note that
$\kappa^{\rm LO}(S,x_1,x_2)$ is finite, $M$-independent function of
these variables.  In the above equations
the factorization scale $M$ is a free parameter, which
separates short distances (large $\mid t_1\mid, \mid t_2\mid$
of the incoming partons in Fig.3) from large distances,
i.e. small virtualities, when both incoming partons $a,b$ are close to
their mass shell. The formal invariance of the factorization procedure
with respect to the choice of $M$
is guaranteed by the cancellation
mechanism which, in the above process, works in such a way that the
$M$-dependence of the LO contribution to (\ref{sigmahadrons}), contained
in the $M$-dependent distribution functions $D_{a/p}(x,M)$
and described by Feynman diagrams of Fig.3a, is cancelled to the NLO
by the explicit dependence on $M$ of $\kappa^{\rm NLO}$,
corresponding to diagrams of Fig.3b.
\footnote{Similar procedure can be formulated for the case of final
state collinear singularities. These are either put into the appropriate
fragmentation functions, or cancelled, via KLN theorem, in the sum over
all indistinguishable partonic final states.}
This cancellation mechanism
works separately for any two-parton final state in the subprocess
(\ref{subprocess}).

In the case of the photoproduction of jets (\ref{photoproduction})
the cancellation mechanism
is modified by the presence of the inhomogeneous term
in the evolution equation for quark distribution function
$D_{q/\gamma}(x,M)$. Let us consider, as an example, the two
gluon final state of Fig.4a and, moreover, fix the lower parton leg,
i.e. the parton coming from the target proton, to be quark of a
particular flavour.  The NLO QCD corrections to this subprocess are
described, in the case of real emissions, by diagrams of Fig.4b.  They
cancel that part of the $M$-dependence of the LO diagrams of Fig.4a,
which is induced by the homogeneous part of the evolution equation for
$D_{q/\gamma}(x,M)$, i.e. the one involving the convolutions with the
quark or gluon distribution functions. They, however, don't cancel the
dependence of $D_{q/\gamma}(x,M)$ on $M$ induced by the inhomogeneous
term in (\ref{quark}). This additional dependence on $M$ clearly needs a
``direct'' photon in the initial state and is therefore cancelled by the
``direct'' $\gamma$ contribution, corresponding to the diagram in
Fig.4c. The cross-section of the $2\rightarrow3$ subprocess
$\gamma$+q$\rightarrow$q+g+g has a parallel singularity, arising from
the configuration where the quark-antiquark pair originating from the
incoming photon is parallel to it, which has to be subtracted. Instead
of (\ref{sigmahadrons}) we thus have
\begin{displaymath}
\sigma(\gamma{\rm +p}\rightarrow {\rm 2\ jets})= \sum_{b}\int_0^1{\rm
d}x_2D_{b/p}(x_2,M) \sigma^{hard}_{\gamma b}(S,x_2,M,\mu)
\end{displaymath}
\begin{equation}
+\sum_{a,b}\int_0^1{\rm d}x_1\int_0^1{\rm d}x_2
D_{a/\gamma}(x_1,M)D_{b/p}(x_2,M) \sigma^{hard}_{ab}(S,x_1,x_2,M,\mu)
\label{sigmaphoton}
\end{equation}
where the direct $\gamma$ hard scattering cross-section
\begin{equation}
\sigma^{hard}_{\gamma b}(S,x_2,M,\mu)=
\left[\alpha\alpha_s(\mu)\kappa^{\rm LO}_{\gamma b}(S,x_1,x_2)+\alpha
\alpha_s^2(\mu)\kappa^{\rm NLO}_{\gamma b}(S,x_1,x_2,M,\mu)\right]
\label{hardphoton}
\end{equation}
contains the $M$-independent LO contribution, corresponding to
diagram in Fig.4d, and $M$-dependent NLO part, which cancels the
rest of the $M$-dependence of the LO ``resolved'' $\gamma$ contribution
induced by the inhomogeneous term in (\ref{quark})! Thus the
inclusion of the NLO direct $\gamma$ contribution is neccessary
for a theoretically consistent
description of jet production in $\gamma$p collisions
to the order ${\cal O}(\alpha\alpha_s^3)$.
As far as $\kappa^{\rm NLO}_{\gamma b}$ is concerned,
one has to be careful as this quantity depends on the
choice of the NLO branching function $P^{(1)}(z)$. Moreover, we cannot
just subtract from the integrand of the appropriate Feynman
diagrams the singular part, corresponding to the $1/t$ term,
as the $M$-dependence enters
through the upper limit on the interval of $\mid t\mid$,
where this subtraction is carried out. This question is
discussed in more detail in \cite{ja}.

At the ${\cal O}(\alpha\alpha_s^2)$ order the factorization scale
dependence doesn't cancel, but it is clear that the $M$-dependent NLO
direct $\gamma$ term has to be included along with the LO resolved
$\gamma$ one, as they are both of the same order.
For instance, in a recent analysis \cite{Torsjo}
of the hadronic properties of the photon the total cross-section
of $\gamma$p interactions is written as a sum of three terms, the first
two of them corresponding to the ``hadronic'' and asymptotic pointlike
components of the resolved $\gamma$ contribution, while the third one is
the LO direct $\gamma$ contribution.
However, as emphasized above, any time the LO resolved $\gamma$
contribution is taken into account, so must also be, for theoretical
consistency, the NLO direct $\gamma$ one!
Moreover, as in \cite{Torsjo} the scale $M$ of
$D_{\rm NS}^{ap}(x,M)$ is taken very small ($\approx 0.5$ GeV), the
NLO direct $\gamma$ contribution evaluated at this scale may be
quite large and its final state practically indistinguishable from
the resolved $\gamma$ one. Although including this NLO direct $\gamma$
component means adding
another positive contribution and thus further complicating
the situation in \cite{Torsjo}, there is no justification for
neglecting it.

Finally, at the ${\cal O}(\alpha\alpha_s)$ order, only the
$M$-independent direct
$\gamma$ component contributes, which has no parallel singularity
once the minimal transverse momentum of the produced
jets is specified. The q$\overline{{\rm q}}$ pair originating from the
photon in Fig.4d can not be parallel and still produce nonzero $p_{t}$
at the lower vertex.

In the case of $W$ boson production in $\gamma$p collisions (\ref{W})
the cancellation mechanism starts earlier
than in the preceding subcase. The lowest order
diagrams corresponding to (\ref{W}) are, for direct as well as resolved
$\gamma$ components, sketched in Figs.5a,c.
Both of these contributions
are of the order ${\cal O}(\alpha\alpha_W)$.
   As the produced $W$ is massive,
the q$\overline{\rm q}$ pair originating in Fig.5c
                                         from the incoming
photon can, contrary to diagram in  Fig.4d, be parallel.
Consequently, the parallel
singularity of the ${\cal O}(\alpha\alpha_W)$ direct $\gamma$ term
has to be subtracted, leading to the $M$-dependent finite part. The
subtracted, $M$-dependent, part is included in the
$M$-dependent photon distribution function
$D_{q/\gamma}(x,M)$,  entering
the LO contribution to the resolved $\gamma$ component of Fig.5a.
Higher order QCD corrections (Fig.5b) work as before.
In this case the mechanism of factorization thus starts to
operate already at the order ${\cal O}(\alpha\alpha_W)$ and
provides another evidence against
the claim that $D_{q/\gamma}(x,M)={\cal O}(\alpha/\alpha_s)$. Used
in this process the latter behaviour
 would imply for the resolved component
$\sigma^{res}(\gamma +{\rm p}\rightarrow W)=
         {\cal O}(\alpha\alpha_W/\alpha_s)$.
Consequently, the LO direct and resolved contributions would be of
different orders in $\alpha_s$ and, worst of all, $\sigma^{res}$
would diverge for $\alpha_s\rightarrow0$.

So far I have discussed only the real photon interactions.
In ep collisions the photon exchanged
in NC interactions ($Z^0$ exchange can be neglected in the region
of $Q^2$ considered) is virtual and one has to be careful to treat
properly the dependence of photon distribution functions
$D_{i/\gamma}(x,M,Q^2)$, $i$=q, $\overline{\rm q}$, g
                                                   on the virtuality
$Q^2$ of the exchanged photon \cite{Schuler}. In this note I
restrict my attention to the region of low $Q^2$, where the virtual
photon behaves to a very good approximation
            like the real one. The spectrum of
these photons inside the incoming electron is given by the
Weizs\"{a}cker-Williams approximation
\begin{equation}
D_{\gamma/e}(x,Q_0)=\frac{\alpha}{2\pi}\left(\frac{1+(1-x)^2}{x}\right)
\ln\frac{Q^2_0(1-x)}{m_{\rm e}^2x^2},
\label{WW}
\end{equation}
where $Q_0$ denotes
the maximum virtuality of the exchanged $\gamma$ taken into
account. The ``optimal'' choice of $Q_0$ (in the sense that the
cross-section of the 2$\rightarrow$3 subprocess
e+q/g$\rightarrow$e+q+g/$\overline{\rm q}$ behaves as 1/$Q^2$ up
to $Q^2=Q_0^2$) depends on the transverse
momentum of produced partons and is roughly linear function thereof.
My considerations concern this region.

Up to now all theoretical analyses of data on jet production in
ep collisions have included, along with the resolved
$\gamma$ contribution of Fig.4a, also (and only) the LO
direct $\gamma$ one of Fig.4d. These analyses show that around
$p_t^{c}\approx25$ GeV,
${\rm d}\sigma^{res}/{\rm d}p_t\doteq{\rm d}
\sigma^{dir}/{\rm d}p_t$.  As emphasized above,
the ${\cal O}(\alpha^2\alpha_s^2)$ resolved $\gamma$ component should,
however, always be considered simultaneously with NLO direct $\gamma$,
which is of the same order.
Although there exists \cite{Boedeker} a comprehensive analysis
of direct $\gamma$ contributions to jet production in ep collisions
up to the NLO, there is so far no theoretically consistent
treatment of full jet production to the order
${\cal O}(\alpha^2\alpha_s^2)$, including LO resolved (Fig.4a)
                    together with
LO+NLO direct $\gamma$ contributions (Fig.4d,c).
The results of \cite{Bod2} show that the ratio
$r\equiv
        ({\rm d}\sigma^{\rm NLO}/{\rm d}p_t)/
             ({\rm d}\sigma^{\rm LO}/{\rm d}p_t)$
is of the order of 2 in the region of $p_t$ around 20 GeV and that
$r$ grows as $p_t$
decreases so that the inclusion of NLO direct $\gamma$ contribution
shifts the crossing point $p_t^{c}$ to smaller values. It would be
very useful to have the calculations of \cite{Boedeker,Bod2}
available down to $p_t$ around 10 GeV,
where most of the available HERA data are, to see how much the
situation will actually change by including the NLO direct $\gamma$
term.

Moreover, in these circumstances
the final state in the NLO direct $\gamma$ channel will often
lead to kinematically very close final state configurations
as the resolved one. This is due to the fact that
the on-mass shell quark/antiquark, originating from the incoming
photon and not participating in further
interaction with the constituents of the proton, may fly
close to the direction of the parent photon and
thus is essentially distinguishable from the ``remnant'' jets of the
resolved $\gamma$, Fig.4a.
For small $p_t$ the factorization mass $M$ to be used in
$D_{q/\gamma}(x,M)$ should be roughly proportional to $p_t$,
which implies small opening angle between q$\overline{\rm q}$
in the the direct $\gamma$ contribution of Fig.4c.
                For high $p_t$ the ``remnant''
jet from NLO direct component is expected to have bigger angle
with respect to the beam ($\gamma$) direction and thus to be
distinguishable from the resolved $\gamma$ one.

Let us now return to the interpretation of
recent HERA data on two jet events. In \cite{H1,Zeus}
                    only the LO direct and LO resolved
$\gamma$ contributions were included. As
theoretically consistent ${\cal O}(\alpha^2\alpha_s^2)$ order
analysis requires the inclusion
of the NLO direct $\gamma$ term, the interpretation of the
mentioned observation as an evidence for resolved $\gamma$
component is premature.
What can be safely said is that the data require the presence
of the ${\cal O}(\alpha^2\alpha_s^2)$ effects. Even this, however,
is an important conclusion.

\vspace{0.4cm}
\parindent 0.001cm
{\Large {\bf Acknowledgment}}

\vspace{0.3cm}
I am grateful to P. Kol\'{a}\v{r} for careful reading of the manuscript
and numerous critical comments and suggestions.

\newpage
\vspace{0.3cm}
\parindent 0.01cm
{\Large \bf Figure captions} \\

\vspace{0.3cm}
Fig.1: Basic QED vertex (a) and example of lowest order QCD correction
to it (b). In this and all following figures
the big solid circles stand for parton distribution functions of the
colliding hadrons (labelled by double lines) or the photon.

\vspace{0.3cm}
Fig.2: Examples of Feynman diagrams contributing to the resolved (a)
and direct (b) $\gamma$ photoproduction of jets.

\vspace{0.3cm}
Fig.3: Factorization for the two gluon final
state in proton-proton collisions:
  the $M$-dependence of $D_{q/{\rm p}}(x,M)$ used in
${\cal O}(\alpha_s^2)$ Feynman diagram of
   (a) is compensated by the explicit $M$-dependence of the
${\cal O}(\alpha_s^3)$ hard scattering cross-section (b).

\vspace{0.3cm}
Fig.4: Factorization for the two gluon final state in
$\gamma$p collisions:
  the $M$-dependence of $D_{q/\gamma}(x,M)$ used in
${\cal O}(\alpha\alpha_s^2)$ Feynman diagram for resolved $\gamma$
  (a) is compensated in part by the explicit $M$-dependence
 of the ${\cal O}(\alpha\alpha_s^3)$ resolved $\gamma$ hard scattering
 cross-sections (b) and in part
by the $M$-dependence of the ${\cal O}(\alpha\alpha_s^2)$
direct $\gamma$ hard scattering cross-section (c).
In (d) the LO, ${\cal O}(\alpha\alpha_s)$, direct $\gamma$
                       contribution is plotted for comparison.

\vspace{0.3cm}
Fig.5: Factorization in photoproduction of $W$ boson:
  the $M$-dependence of $D_{q/\gamma}(x,M)$ used in ${\cal O}(\alpha
\alpha_W)$ resolved $\gamma$ contribution (a) is cancelled
                                                in part by the
explicit $M$-dependence of ${\cal O}(\alpha\alpha_W\alpha_s)$ hard
 scattering cross-sections (b) and
in part by the explicit $M$-dependence of ${\cal O}(\alpha\alpha_W)$
hard scattering cross-section (c). Dashed lines denote $W$ boson.
\end{document}